\documentclass[aps,prl,twocolumn,groupedaddress,showpacs]{revtex4-1}
\usepackage{graphicx}
\usepackage{psfrag}
\def\be{\begin{equation}}
\def\ee{\end{equation}}

\begin{document}
\date{\today}

\title{Non-Equilibrium Phase Transitions in Systems with Long-Range Interactions}
\author{Tarc\'isio N. Teles}
\author{Fernanda Benetti}
\author{Renato Pakter}
\author{Yan Levin}
\affiliation{Instituto de F\'{\i}sica, Universidade Federal do Rio Grande do Sul, Caixa Postal 15051, CEP 91501-970,
Porto Alegre, RS, Brazil}

\begin{abstract}

We introduce a generalized Hamiltonian Mean Field Model (gHMF) --- XY model with both linear and quadratic coupling between spins and explicit Hamiltonian dynamics.   
In addition to the usual paramagnetic 
and ferromagnetic phases, this model also possesses a nematic phase.  
The gHMF can be solved explicitly using
Boltzmann-Gibbs (BG) statistical mechanics, in 
both canonical and
microcanonical ensembles.  However, when the resulting {\it microcanonical} phase diagram 
is compared with  the one obtained using molecular dynamics simulations, it is found  that the two are very different.  We will present a dynamical theory which allows us to explicitly calculate the phase diagram obtained
using molecular dynamics simulations without any adjustable parameters.
The model illustrates the fundamental role played by dynamics as well the inadequacy of BG statistics for systems with long-range forces {\it in the thermodynamic limit}.

\end{abstract}


\maketitle

A fundamental concept in statistical mechanics, taught in a typical course, is
equivalence of ensembles~\cite{Hua1987}.
One also learns that mean-field theory becomes exact for systems with long-range (LR) interactions~\cite{Kac1963,Sta1987}.
However, in order to have a well 
defined thermodynamic limit, in this case, 
special care must be taken. The usual
approach is to scale the strength of the two-body interaction potential with the number of particles
in the system, $N$.  This is the, so-called, Kac prescription --- it makes the infinitely-long-range two-body interaction infinitesimally weak~\cite{Kac1963}.  The thermodynamic limit becomes well defined, since both the kinetic and the potential contributions to the total energy now scale linearly with $N$, making the energy extensive.  
Over the last decade, however, it has
become clear that both the ensemble equivalence and the exactness of mean-field theory
fail for systems with LR interactions~\cite{Barre2001,Campa09,FilAma2009}.  
The phase-diagrams calculated using Boltzmann-Gibbs (BG) statistics
in canonical and microcanonical ensembles do not always coincide~\cite{Barre2001}. 
Furthermore, molecular dynamics simulations, 
show that  isolated 
LR interacting systems become
trapped in quasi-stationary states (qSS) the life time of which diverges with the number of particles~\cite{LatRap1999,YamBar2004,AnFa07,Kav2007,JoWo,Levinprl,TaLe10,Tel2011,PaLe11}. 

The inapplicability of BG statistics to systems with LR forces in thermodynamic limit 
is a consequence of  
the ergodicity breaking. 
Scaling of two-body potentials with the number of particles --- essential
for the existence of a well defined thermodynamic limit --- destroys the 
correlations (collisions) between the particles~\cite{Levinrep} that drive normal short-range interacting systems towards the thermodynamic equilibrium.
Relaxation to the stationary state of an LR system
is, therefore, fundamentally different from the collisional (correlational)
relaxation of normal gases and fluids.  Collisionless relaxation  
relies on the collective excitations and evaporative cooling driven by 
Landau damping~\cite{La46,Levinprl}.  The final stationary state reached by a collisionless system  
is intrinsically non-ergodic~\cite{TaLe10,BeTe2012}. 
It does not correspond to the maximum of the
Boltzmann entropy.  
To exemplify this
dichotomy, in this Letter we introduce a new generalized Hamiltonian Mean Field Model (gHMF)  --- a  LR version of
the model studied in ref.~\cite{PoAr2011,LeGr1985}  --- 
which can be solved exactly using BG statistical mechanics.  We will
show that the equilibrium phase diagram predicted by the BG statistics
in the microcanonical ensemble is
very different from the one obtained using the 
molecular dynamics (MD) simulations.  We will then
construct a dynamical theory that correctly predicts the location and the 
order of the phase transitions
observed in MD simulations.

The gHMF is described by the Hamiltonian 
\begin{eqnarray}
\label{eq:ham}
H(\theta_i,p_i)= &\sum_{i=1}^{N}\frac{p_i^2}{2}+\frac{1}{2N}\sum_{i,j=1}^{N} \left[1-\Delta\cos(\theta_i-\theta_j)\right.\nonumber\\
\, &\left.-(1-\Delta)\cos(2 \theta_i -2 \theta_j)\right],
\end{eqnarray}
where $\Delta \in [0,1]$.  
The model can be thought of as either XY-spins confined to
a line, or as particles restricted to move on a circle.  The latter interpretation is
perhaps more convenient when discussing MD simulations with equations of motion given by: 
$\dot{\theta_i}=\partial H/\partial p_i$ and $\dot{p_i}=-\partial H/\partial \theta_i$.

We define the ferromagnetic and nematic order parameters as 
$m_1=  \frac{1}{N}\sum_{i=1}^{N}\cos\theta_i$ and 
$m_2=  \frac{1}{N}\sum_{i=1}^{N}\cos 2\theta_i$,
respectively.
Using the usual statistical mechanics approach~\cite{Campa09}, we first calculate the microcanonical entropy for the gHMF. 

\begin{figure}
\begin{center}
\vspace{0.5cm}
\includegraphics[scale=0.7,width=7.5cm]{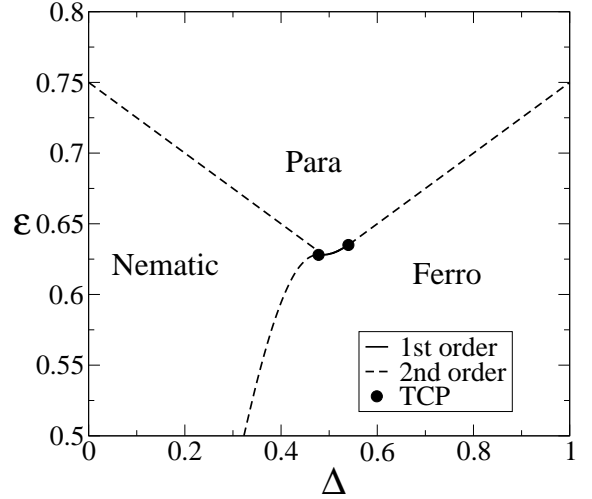}
\caption{Microcanonical phase diagram obtained using BG statistics. Solid circles
are the two tricritical points.\label{fig:mcq2}}
\end{center}
\end{figure}

Within BG statistical mechanics all the thermodynamic information is encoded in the phase
space volume accessible to the system with the total energy $E$
\begin{equation}
\label{eq:microensemble}
\Omega(E,N)=\int_{-\pi}^{\pi}\prod d\theta_i \int_{-\infty}^{\infty}\prod dp_i \delta(E-H({\theta_i},{p_i})).
\end{equation}
The integral in Eq.  (\ref{eq:microensemble}) can be divided into two parts -- kinetic and configurational,
\begin{equation}
\Omega(E,N)=\int dK \Omega_{kin}(K)\Omega_{conf}(E-K)
\end{equation}
where
\begin{eqnarray}
\Omega_{kin}(K)&=&\int_{-\infty}^{\infty}\prod dp_i \delta\left(K-\frac{\sum p_i^2}{2}\right),\\
\Omega_{conf}(E-K)&=&\int_{-\pi}^{\pi}\prod d\theta_i  \delta(E-K-U(\{\theta_i\})),
\end{eqnarray}
and $U$ is the potential energy, second term in Eq. (\ref{eq:ham}).
Integrating over the momentum degrees of freedom, in the thermodynamic limit we obtain
\begin{equation}
\Omega_{kin}(K)=\exp\left[\frac{N}{2}\left(\ln\pi+\ln 2K-\ln \frac{N}{2}+1\right)\right],
\end{equation}
The microcanonical entropy per particle is
$s(\varepsilon)=\frac{1}{N}\ln\Omega(E,N)$ 
\begin{equation}
\label{eq:mcentropy1}
s(\varepsilon)=\frac{1}{2}\ln 2\pi +\frac{1}{2} + \sup_{\kappa}\left[\frac{1}{2}\ln 2\kappa + \frac{1}{N}\ln \Omega_{conf}(N(\varepsilon-\kappa))\right].
\end{equation}
where $\kappa \equiv K/N=(E-U)/N=\varepsilon-u$.  
Since the potential energy depends only on $m_1$ and $m_2$, we define
\begin{eqnarray}
\Omega_m(m_1,m_2)=&\int_{-\pi}^{\pi}\prod d\theta_i  \delta(\sum\cos\theta_i-Nm_1)\nonumber \\
\,&\times \delta(\sum\cos 2\theta_i-Nm_2),
\end{eqnarray}
which using the Fourier representation of the delta function can be written as, 
\begin{eqnarray}
\Omega_m &(m_1,m_2)=\frac{1}{(2\pi)^2}\int_{-\infty}^{\infty}dx\int_{-\infty}^{\infty}dy
\exp \left\lbrace N\left[-ixm_1\right.\right.  \nonumber \\
&\left.\left.-iym_2+\ln\left(\int d\theta \exp(ix\cos\theta +iy\cos 2\theta\right)\right]\right\rbrace.
\end{eqnarray}
The integral can be evaluated using the saddle-point method. The extremum corresponds to $(x^{\star},y^{\star})$, which must satisfy
\begin{eqnarray}
m_1&=\frac{\int d\theta \cos\theta \exp\left[ix\cos\theta +iy\cos 2\theta\right]}{\int d\theta \exp\left[ix\cos\theta +iy\cos 2\theta\right]},\label{eq:xcondition}\\
m_2&=\frac{\int d\theta \cos 2\theta \exp\left[ix\cos\theta +iy\cos 2\theta\right]}{\int d\theta \exp\left[ix\cos\theta +iy\cos 2\theta\right]}.\label{eq:ycondition}
\end{eqnarray}
Defining $a=ix^{\star}$ and $b=iy^{\star}$ and neglecting terms of order lower than $N$,
\begin{eqnarray}
&\frac{1}{N}\ln \Omega_{m}(m_1,m_2)=-m_1 a(m_1,m_2)-m_2 b(m_1,m_2) \nonumber \\
& + \ln\left(\int d\theta \exp\left[a(m_1,m_2)\cos\theta +b(m_1,m_2)\cos 2\theta\right]\right).
\end{eqnarray}
In the thermodynamic limit, we may replace $\ln \Omega_{conf}(E-K)$ by $\ln \Omega_{m}(m_1,m_2)$ in Eq. \ref{eq:mcentropy1}.  Furthermore, noting that $\kappa = \varepsilon - u$, where $u=(1-\Delta m_1^2- (1-\Delta) m_2^2)/2$, the maximization can be taken with respect to $m_1,m_2$ instead of $\kappa$. The entropy per particle then becomes
\begin{eqnarray}
\label{eq:mcentropyfinal}
&s(\varepsilon)=\frac{1}{2}\ln 2\pi +\frac{1}{2}+ \sup_{m_1,m_2}\left[\frac{1}{2}\ln\left(2\varepsilon -1+\Delta m_1^2\right.\right. \nonumber \\
&\left.+(1-\Delta) m_2^2\right)-m_1 a(m_1,m_2)-m_2 b(m_1,m_2) \nonumber \\
&\left.+\ln \left(\int d\theta \exp[a(m_1,m_2)\cos\theta +b(m_1,m_2)\cos 2\theta]\right)\right].
\end{eqnarray}
with the equilibrium values of the order parameter $(m_1^{\star},m_2^{\star})$ given by 
\begin{eqnarray}
\label{eq:mcmagsolution}
\frac{\Delta m_1^{\star}}{2\varepsilon -1+\Delta m_1^{\star 2}+(1-\Delta)m_2^{\star 2}}=a(m_1^{\star},m_2^{\star}), \\
\frac{(1-\Delta)m_2^{\star}}{2\varepsilon -1+\Delta m_1^{\star 2}+(1-\Delta) m_2^{\star 2}}=b(m_1^{\star},m_2^{\star}).
\end{eqnarray}
Substituting these expressions into Eqs. (\ref{eq:xcondition}) and (\ref{eq:ycondition}),
we find the equilibrium values of the order parameters
\begin{eqnarray}
\label{eq:magsolveMC}
m_1=& \frac{\int_{-\pi}^{\pi}d\theta\cos\theta\exp\left[\frac{\Delta m_1\cos\theta+(1-\Delta)m_2\cos 2\theta}{2\varepsilon -1+\Delta m_1^2+(1-\Delta)m_2^2}\right]}{\int_{-\pi}^{\pi}d\theta\exp\left[\frac{\Delta
m_1\cos\theta+(1-\Delta)m_2\cos 2\theta}{2\varepsilon -1+\Delta m_1^2+(1-\Delta)m_2^2}\right]},\\
m_2=& \frac{\int_{-\pi}^{\pi}d\theta\cos 2\theta\exp\left[\frac{\Delta m_1\cos\theta+(1-\Delta)m_2\cos 2\theta}{2\varepsilon -1+\Delta m_1^2+(1-\Delta)m_2^2}\right]}{\int_{-\pi}^{\pi}d\theta\exp\left[\frac{\Delta
m_1\cos\theta+(1-\Delta)m_2\cos 2\theta}{2\varepsilon -1+\Delta m_1^2+(1-\Delta)m_2^2}\right]} \label{eq:magsolveMC2},
\end{eqnarray}
where for notational simplicity we have dropped $^{\star}$.
In the case of a first order phase 
transition --- more than one solution of  Eqs.(\ref{eq:magsolveMC}) and (\ref{eq:magsolveMC2})--- the equilibrium  values of $m_1$ and $m_2$ will correspond to the ones that lead to the maximum  entropy.
The resulting microcanonical phase diagram is shown in Fig. \ref{fig:mcq2}. 

\begin{figure}
\vspace{0.5cm}
\includegraphics[scale=0.7,width=7.5cm]{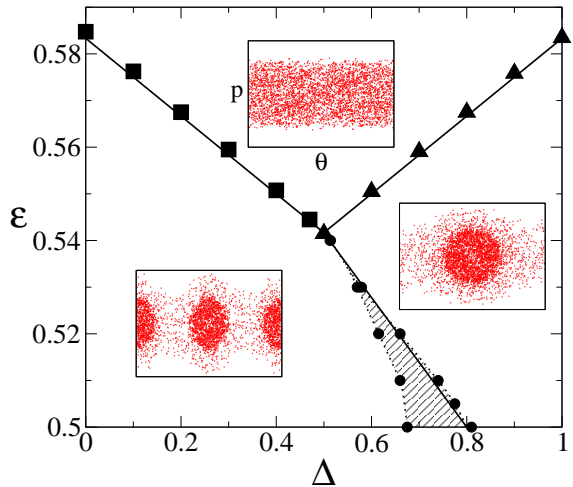} 
\caption{The out-of-equilibrium phase diagram of the gHMF. The squares and triangles are simulation results for the qSS nematic-paramagnetic and para-ferromagnetic phase transitions, respectively. The shaded area represents the nematic-ferromagnetic transition region in which either phase occurs with equal probability. To the right of this region, the order is ferromagnetic, and to the left, nematic.  Black solid lines are the theoretical predictions for the transitions. All transitions are first order.
Insets show the phase space particle distribution in different phases.  Notice the
characteristic core-halo structure~\cite{PaLe11} both inside nematic and ferromagnetic phases. The simulations were performed with $N=10^6$ particles for the paramagnetic-nematic and paramagnetic-ferromagnetic transition, and with $N=10^7$ particles 
to locate the instability region between the nematic and ferromagnetic phases.\label{fig:qssdiagram}}
\end{figure}

Equation (\ref{eq:microensemble}) requires that the system described by the Hamiltonian
(\ref{eq:ham}) is ergodic --- has equal probability of visiting all possible microstates. To see
if this is the case,  we  use  Molecular Dynamics (MD) simulations to study its dynamics.
For the gHMF, we are interested to understand how an ordered (ferromagnetic or nematic)  state can 
arise from an originally disordered  homogeneous (paramagnetic) particle distribution 
$f_0(\theta,p)={1\over 4\pi p_0} \Theta (\pi-|\theta|)\, \Theta (p_0-|p|)$. 
The Hamilton's equations of motion reduce to a second order differential equation for 
$\theta_i$,
\begin{eqnarray}
\ddot\theta_i &=& F(\theta_i) \label{evol} \\  \nonumber
&\equiv& -\Delta m_1(t)\sin \theta_i(t)-2(1-\Delta) m_2(t)\sin 2 \theta_i(t).
\end{eqnarray}
where $F(\theta)$ is the force acting on a particle located at $\theta$, and 
where we have used the fact that 
$\langle \sin \theta(t) \rangle=\langle \sin 2 \theta(t) \rangle=0$, throughout
the dynamical evolution~\cite{PaLe11,Mar12}.  
Comparing the phase diagram obtained using MD simulations, we see that it
is very different from the prediction of the microcanonical BG statistical mechanics,
see Fig. \ref{fig:qssdiagram}. 

Besides occurring in different regions of the $(\varepsilon,\Delta)$ plane, the 
phase transitions predicted by the BG statistics are of the wrong order! While the transitions
from paramagnetic to ferromagnetic or nematic phases are
found to be of second order,  MD simulations show that these transitions are of
first order.  Furthermore, the second order phase transition line between the nematic and the ferromagnetic
phase disappears completely and is replaced by a region of instability in which
either phase can occur with equal probability.

To understand the results of MD simulations, one must forget equilibrium 
statistical mechanics and return to kinetic theory.
In the thermodynamic limit, the dynamical evolution of the one-particle distribution function
$f(\theta,p,t)$
of a system with long-range interactions is governed exactly by the Vlasov equation~\cite{Br77}. Vlasov
dynamics is collisionless  --- the relaxation to equilibrium comes 
from Landau damping, a dynamical process in which 
individual particles gain energy from collective oscillations, while the oscillations are
damped out.  The one-particle energy of the gHMF is
 $\epsilon=p^2/2+ 1-\Delta m_1 \cos(\theta)-(1-\Delta)m_2 \cos(2 \theta)$.
Note that the initial particle distribution $f_0(\theta,p)$ has $m_1=m_2=0$,  so that it
can be expressed as
a function of $\epsilon$.  This means that $f_0(\theta,p)$ is a 
stationary solution of the
Vlasov equation.  A phase transition in gHMF, therefore, can occur only after a dynamical
instability.  To explore the non-linear stability of the gHMF, we consider a perturbation
of the initial distribution, such that the maximum momentum  $p_0 \rightarrow p_m(t)=p_0+\sum_{n=0}^\infty  A_n(t) \cos(n\theta)$.  We define the generalized order parameters as
\be  
m_n(t) \equiv \langle \cos(n \theta)\rangle \equiv \int f(\theta,p,t) \cos(n\theta)dpd\theta
\label{mn}
\ee 
where $f(\theta,p,t)={1\over 4\pi p_0} \Theta (\pi-|\theta|)\, \Theta (p_m(t)-|p|)$. Note
that this distribution preserves the phase space density, as is required by the Vlasov equation.  Performing the integration in Eq. (\ref{mn}), we find that $m_n(t)=A_n(T)/2 p_0$.  Taking
two temporal derivatives of $m_n(t)$, we obtain,
\be
\ddot m_{n}=-n^2 \langle p^2\cos(n\theta) \rangle-n\langle 
F(\theta)\sin(n\theta) \rangle ,
\label{mpp}
\ee
where we have used the equation of motion, Eq.(\ref{evol}).  Performing the averages
using the distribution function $f(\theta,p,t)$, we obtain the equations of motion for
the generalized order parameters,
\begin{eqnarray}
\ddot m_1+\left( {12 \varepsilon-6-\Delta\over 2} \right )m_1 = f_1(m_1,m_2,m_3,m_4) \label{op1} \\
\ddot m_2+2\left( 12 \varepsilon+\Delta-7 \right )m_2 = f_2(m_1,m_2,m_3,m_4) \label{op2}\\
\ddot m_3+ 27 (2 \varepsilon-1) \, m_3 = f_3(m_1,m_2,m_3,m_4) \label{op3}\\
\ddot m_4+ 48 (2 \varepsilon-1) \, m_4 = f_4(m_1,m_2,m_3,m_4) \label{op4}
\end{eqnarray}
where 
\begin{widetext}
\begin{eqnarray}
f_1&=&{m_1}\,m_2\left( 1 - \frac{3\,\Delta}{2}
     \right)  + \left(\Delta-1  \right) \,{m_2}\,
   {m_3}-\nonumber\\ &3&\left( 2\,\varepsilon -1\right) \,
   \left\{ {{m_1^3}} + {{m_1^2}}\,{m_3} + 
     {m_3}\,\left[ {m_2}\,\left( 2 + {m_2} \right)  + 
        2\,\left( 1 + {m_2} \right) \,{m_4} \right]  + 
     2\,{m_1}\,\left[ {m_2} + {{m_2^2}} + {{m_3^2}} + 
        {m_2}\,{m_4} + {{m_4^2}} \right]  \right\}\\
 f_2&=& \Delta \,
   \left( {{m_1^2}} - {m_1}\,{m_3} + 
     2\,{m_2}\,{m_4} \right)-2\,{m_2}\,{m_4}-\nonumber\\  &12& \left( 2\,\varepsilon-1 \right) \,
   \left[ {{m_2^3}} + {{m_3^2}}\,{m_4} + 
     2\,{m_1}\,{m_3}\,
      \left( 1 + {m_2} + {m_4} \right)  + 
     {{m_1^2}}\,\left( 1 + 2\,{m_2} + {m_4} \right)  + 
     2\,{m_2}\,\left( {{m_3^2}} + {m_4} + {{m_4}^2}
        \right)  \right]\\
  f_3 &=& \frac{3\,{m_1}}{2}\,\left[ \left( 2 - \Delta  \right) \,
        {m_2} - \Delta \,{m_4} \right] -\nonumber \\ &9&\left( 2\,\varepsilon -1\right)
   \left\{ {{m_1^3}} + 6\,{{m_1^2}}\,{m_3} + 
     3\,{m_1}\left[ {m_2}
         \left( 2 + {m_2} \right)  + 
        2\,\left( 1 + {m_2} \right) \,{m_4} \right]  + 
     3\,{m_3}\left[ {{m_3^2}} + 
        2\,\left( {{m_2^2}} + {m_2}\,{m_4} + 
           {{m_4^2}} \right)  \right]  \right\}\\
  f_4 &=& 2 \Delta \, m_1m_3-4(\Delta -1) m_2^2- \nonumber\\
&48& \left(2 \varepsilon -1 \right)  
  \left[ 2 {m_1} \left( 1 + {m_2} \right)  {m_3} + 
    {m_2} \left( {m_2} + {{m_3^2}} \right)  + 
    2 \left( {{m_2^2}} + {{m_3^2}} \right)  {m_4} + 
    {{m_4^3}} + {{m_1^2}} 
     \left( {m_2} + 2 {m_4} \right)  \right]
\end{eqnarray}
\end{widetext}
We have restricted ourselves to the first four generalized order parameters, since
these are already sufficient to understand the phase diagram obtained using MD simulations.
Note that the right hand sides of Eqs. (\ref{op1} -- \ref{op4})
are non-linear functions, so that the transition from paramagnetic-to-ferromagnetic
or paramagnetic-to-nematic phases is determined by the linear stability of  
these equations.  Eqs. (\ref{op1}) and (\ref{op2}) show that the paramagnetic phase  becomes unstable to
ferromagnetic ordering when $12 \varepsilon -6- \Delta <0$ and to nematic ordering when $12 \varepsilon + \Delta -7<0$.  The two stability lines agree perfectly with the results of MD simulations,
see Fig \ref{fig:qssdiagram}.  It is important to note that $m_3$ and $m_4$ always remain
linearly stable (recall that $\varepsilon>0.5$ for the initial distribution).

\begin{figure}
\vspace{0.5cm}
\includegraphics[scale=0.7,width=7.5cm]{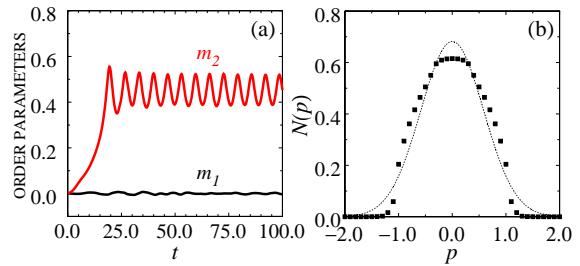} 
\caption{Panel (a) shows the growth and saturation of the order parameter $m_2$ across 
the paramagnetic-nematic transition obtained using MD simulations. The predicted theoretical value is $m_2=0.459$, which is in excellent agreement with the simulations. In panel (b) the symbols are the momentum distribution in the qSS obtained using MD, while the solid line depicts the corresponding Maxwell-Boltzmann distribution 
to which the systems should relax in the infinite time limit. The parameters are 
$\Delta=0.2$ and $u=0.567$. \label{fig3}}
\end{figure}

Linear stability analysis, however, is not sufficient to
determine the order of the phase transitions for which the full non-linear equations
must be considered.  We first note that Eqs. (\ref{op1} -- \ref{op4}) are conservative,  they do not account for the Landau
damping that is responsible for the relaxation to equilibrium and formation of the 
core-halo structures~\cite{PaLe11}, like the ones shown in the insets of  Fig \ref{fig:qssdiagram}.  
Phenomenologically, Landau damping can be included in Eqs. (\ref{op1} -- \ref{op4}) by introducing terms linear in $\dot m_n$.  The relaxation will then proceed
towards the fixed points of Eqs. (\ref{op1} -- \ref{op4}) which can be calculated explicitly.  We find that when either transition line is crossed, the system evolves 
either to nematic $(m_1=0, m_2 \ne 0)$ or ferromagnetic $(m_1 \ne 0, m_2 \ne 0)$ fixed
points.  When crossing the paramagnetic-nematic phase 
transition line, $(\Delta<0.5)$, the order parameter $m_1$ remains zero, while $m_2$
jumps by $\sqrt{\frac{5 \sqrt{43}}{18}-\frac{29}{18}} \approx 0.459 $,  independent of $\Delta$.  This theoretical prediction 
is in excellent agreement with the results of MD simulation which see a jump in the
nematic order parameter of $0.45$, characterizing a strong first-order phase transition, see Fig  \ref{fig3}.
When the paramagnetic-ferromagnetic line is crossed $(\Delta>0.5)$, both $m_1$ and 
$m_2$ experience a jump.  For $\Delta=0.6$, the theory predicts the jumps to be $0.5102$ and  $-0.1861$, for the ferromagnetic and nematic parameters, respectively;
while the simulations find $0.41$ and $-0.10$.   For $\Delta=1$
the theory predicts the respective jumps  to be 
$0.555391$ and $-0.1129$,  while the simulations find $0.45$ and $-0.07$.
It is interesting to note that while for the 
nematic transition
the jump in $m_2$ is universal --- independent of $\Delta$ ---
for the ferromagnetic transition this is not the case. 

What will determine the transition between nematic and ferromagnetic phases?
Deep inside the nematic and ferromagnetic phases, 
Eqs. (\ref{op1} -- \ref{op4}) possess both stable nematic $(m_1=0, m_2 \ne 0)$ and ferromagnetic 
fixed points $(m_1 \ne 0, m_2 \ne 0)$.  
Which of these fixed points is reached first will depend on the initial
condition.  Starting from a paramagnetic distribution $f_0$, in the unstable region
of the phase diagram, both $m_1$ and $m_2$ will grow with time. 
Eqs. (\ref{op1},\ref{op2}) show that the rate of growth of the 
two order parameters are in general very different,  while
$m_1 \sim e^{\lambda_1 t}$, where $\lambda_1=\sqrt{(6+\Delta-12 \varepsilon)/2}$, 
the nematic order
parameter grows as
$m_2 \sim e^{\lambda_2 t}$, with $\lambda_2=\sqrt{14-24 \varepsilon-2 \Delta}$. If the nematic order
parameter first reaches the value characteristic of the nematic fixed point, then
nematic order will be established, otherwise the phase will be ferromagnetic.  
Therefore, we expect that the nematic-ferromagnetic transition line should be given
by $\lambda_1=\lambda_2$ (solid line between nematic and ferromagnetic phases in Fig. \ref{fig:qssdiagram}). This is indeed where the instability characterizing 
nematic-to-ferromagnetic region is found to be, see Fig. \ref{fig:qssdiagram}.

We have introduced a generalized Hamiltonian Mean Field model.  In addition to the
usual paramagnetic and ferromagnetic phases, this model also possesses a nematic phase.
We have obtained the phase diagram of the gHMF using three different methods:  BG statistical mechanics, MD simulations, and a new dynamical theory introduced in 
this paper.  
The model exemplifies the failure of BG statistics
to describe isolated systems with LR interactions, {\it in the thermodynamic limit}.  
This is the first time that a complex (multi-phase)  out-of-equilibrium phase diagram for quasi-stationary states 
has been calculated analytically for a system with LR interactions.

This work was partially supported by the CNPq, FAPERGS, INCT-FCx, and by the US-AFOSR under the grant FA9550-12-1-0438.

\end{document}